\documentclass[letter]{aa}  

\usepackage{graphicx}

\usepackage{txfonts}
\usepackage{ulem}
\usepackage{hyperref}
\begin{document}

\newcommand{\engg}{\mathrm{eng}}
\newcommand{\jett}{\mathrm{jet}}
\newcommand{\ejj}{\mathrm{ej}}

\newcommand{\Int}{\int\limits}
\newcommand\myatop[2]{{{#1}\atop#2}} % "wrapper macro"
\newcommand{\fakeimage}{{\fboxsep=-\fboxrule\fbox{\rule{0pt}{3cm}\hspace{4cm}}}}
\newcommand{\vdag}{(v)^\dagger}
\newcommand\aastex{AAS\TeX}
\newcommand\latex{La\TeX}
\newcommand{\bhac}{\texttt{BHAC}~}
\newcommand{\an}[1]{\textcolor{magenta}{AN: #1}}
\newcommand{\vm}[1]{\textcolor{green}{VM: #1}}
\newcommand{\crr}[1]{\textcolor{red}{CITATION: #1}}
\newcommand{\eg}{e.g.,~}
\newcommand{\ie}{i.e.,~}

\title{Disk mass after a binary neutron star merger  as a constraining  parameter for  short Gamma Ray Bursts}
%\subtitle{BNS merger candidates}
\author{Vasilis Mpisketzis
          \inst{1,2}
          \and
          Antonios Nathanail
          \inst{3}
          }

\institute{Institut für Theoretische Physik, Goethe Universität Frankfurt, Max-von-Laue-Str.1, 60438 Frankfurt am Main, Germany
                \email{vmpisketzis@itp.uni-frankfurt.de}
            \and
            Department of Physics, National and Kapodistrian University of Athens, Panepistimiopolis, GR 15783 Zografos, Greece
            \and
            Research Center for Astronomy, Academy of Athens, Soranou Efessiou 4, GR-11527 Athens, Greece
                \email{anathanail@academyofathens.gr}
            }

\date{December 2023}
\abstract
{The coincident detection of GW170817 and GRB170817A marked a milestone for the
connection between binary neutron star (BNS) mergers and short gamma-ray bursts
(sGRBs). These mergers can lead to the formation of a black hole surrounded by a
disk and the generation of a powerful jet. It spends energy to break free from the merger ejecta, and then a portion of it, is dissipated to produce observable emissions.} 
{ Our primary goal is to enhance our comprehension of
BNS mergers by constraining the disk mass for a selection of
sGRBs, utilizing isotropic gamma-ray luminosity and
corresponding emission times as key indicators.}
{ In this study, we
leverage data from GW170817 to estimate the disk mass surrounding
the BNS merger remnant and subsequently infer the accretion-to-jet
efficiency. Then statistically examine other sGRBs observations to estimate the possibility  of being induced  by BNS mergers}
{Our findings
suggest that, when employing similar physical parameters as in the sole observed BNS-powered GRB event, GRB170817A, a
substantial fraction of sGRBs necessitate an unrealistically
massive disk remnant.}{This observation raises the possibility
that either a different mechanism powered those events or that
the post-collapse disk efficiency exhibits significant variations
across different BNS merger scenarios.}

%\keywords{gravitational waves, gamma-ray bursts}


\maketitle
%\titlerunning
%\authorrunning
%-----------------------------------------------------------------------
\section{Introduction} 
\label{sec:intro}
%-----------------------------------------------------------------------
%Gamma-ray bursts (GRBs) are intense, brief flashes of gamma radiation
%that originate from the distant Universe. %They are some of the most
%energetic events known, releasing more energy in a few seconds than the
%Sun will release in its entire lifetime.  GRBs are divided into two
%classes: long-duration GRBs, which last for more than two seconds, and
%short-duration GRBs, which last for less than two seconds. While
%long-duration GRBs are associated with the collapse of massive stars,
%the origin of short-duration GRBs have the most favorable scenario to be
%a merger of two compact objects.

The detection of  GRB170817A \citep{Goldstein2017}, which was observed
simultaneously with the gravitational wave event GW170817, provided the
first direct evidence that, at least a subset of, sGRBs are
produced by the merger of two neutron stars \citep{2017PhRvL.119p1101A}.
The identification of an electromagnetic optical counterpart to GW170817
\citep{Coulter2017, Arcavi2017, 2017ApJ...848L..32M, 2017Sci...358.1565E}
provided useful insight with respect to open problems in cosmology
\citep{2017Natur.551...85A} and the production of heavy elements
\citep{2017ApJ...848L..27T} but also pinpointed the host galaxy of the
event, allowing for a long-term, multi-wavelength monitoring of the
evolution of the event. This uncovered an additional non-thermal
counterpart that was eventually established as being the afterglow of an
off-axis relativistic jet \citep{Mooley2018,Ghirlanda2019}.

The dynamics of a binary neutron star merger that lead to a
short-duration GRB are strongly affected by the merger process
\citep{Giacomazzo2013}. The two neutron stars spiral together, emitting
gravitational waves.  As they approach each other, they are tidally
deformed. This
tidal deformation leads to the ejection of matter from the system, which
can produce a short-lived, bright electromagnetic transient known as a
kilonova \citep{1998ApJ...507L..59L,2010MNRAS.406.2650M}.  After the
violent merger and the  dynamical ejection of mass
\citep{2015PhRvD..91f4059S, Bovard2017, Radice_2018}, secular mechanisms
further eject mass  through magnetic- and neutrino-driven winds from the
accretion disk and the remnant before its eventual collapse to a black
hole \citep[][and references
therein]{2018ApJ...860...64F,2019ApJ...876..139G,2021JPlPh..87a8402A}

During the  merger  process  a massive, hot accretion disk is produced
around the central object.  Assuming a black hole is formed, accretion of
matter onto it can power the production of a relativistic jet
\citep{Blandford1977, Cruz2022}. The jet then propagates through the
surrounding ejecta, and if it can break out from the ejecta, dissipation
of energy in radiation can be observed \citep{rees:94, McKinney2012b,
Zhang2011c}.  The subsequent interaction of the jet with the surrounding
medium can give rise to afterglow emission, which is observed at longer
wavelengths after the prompt gamma-ray emission has faded
\citep{1992MNRAS.258P..41R}.

Multi-messenger observation of GRB170817A, allowed us to estimate the kinetic
energy of the jet, to be approximately $E_{k,jet}\approx 10^{50}\mathrm{erg}$.
A combination of observables indicated that the remnant collapsed to a black
hole in $t_\mathrm{coll} \approx 1\mathrm{s}$ after merger \citep{2019ApJ...876..139G}.  The surrounding
disk mass was estimated within the limits dictated by numerical relativity
simulations $M_\mathrm{disk}>0.04 M_{\odot}$ \citep{2019EPJA...55...50R}.  From the deduced
energetics of the jet and the estimation of the disk mass it was possible to
infer the efficiency of the accretion power into jet energy
\citep{2021A&A...645A..93S}.

In this work, we take into account the continued mass ejection that feeds
the ejecta through the survival time of the BNS merger remnant.  The jet's
propagation through the ejecta is connected with the observed short GRB
parameters ($t_{\mathrm{GRB}},L_{\mathrm{GRB,iso}}$), and allows us to
arrive to a posterior distribution for the disk mass.  Finally, using the
inferred disk mass distribution we compute the probability that  a
specific short GRB comes from a  BNS merger scenario, based on  the value
of the  inferred disk mass.  The main assumption of this work lies on the
use of the posterior distribution for the efficiency of converting the
mass accretion energy to jet energy for GRB170817A
\citep{2021A&A...645A..93S}.

This letter is organized as follows: in section \ref{sec:setup} we describe the way the
observables are combined with theory to produce posterior distributions
for efficiency and eventually the disk mass for short GRB observations.
In section \ref{results} the results are presented and in section \ref{conc} we conclude.

%
%-----------------------------------------------------------------------
\section{Estimation of dynamical quantities} 
\label{sec:setup}
%-----------------------------------------------------------------------

%\subsection{Algorithm for Estimating Dynamical Quantities}

We present the algorithm designed to estimate various
dynamical quantities based on observed parameters, such as the
isotropic-equivalent GRB luminosity ($L_{\mathrm{GRB,iso}}$) and the
burst duration ($T_{90}$). Our goal is to deduce the mass of the
accretion disk surrounding the remnant black hole by applying robust
statistical assumptions.  

To this scope we need to link observables to dynamical quantities. We
initially associate the isotropic-equivalent jet power
$L_{\mathrm{jet,iso}}$ with $L_{\mathrm{GRB,iso}}$, using a fixed
efficiency parameter ($\epsilon _{GRB} = 0.15$), which remains constant
throughout our calculations. It's worth noting that this parameter
typically falls within the range of $10\%$ to $20\%$ in the literature
\citep{Zhang2015}, so even when considering  a
distribution within the standard range, the impact on the results is insignificant. 
To estimate the available amount of jet energy, we take into account two fundamental factors:
Firstly, the mass of the remnant disk that will be accreted during
the collapse process, and secondly, the efficiency of converting the
accretion energy from the infalling matter into jet energy ($\epsilon
_{disk}$).  For each quantity involved in our calculations, we adopt the
following approach. 

\begin{figure*}
        \includegraphics[width=.46\textwidth]{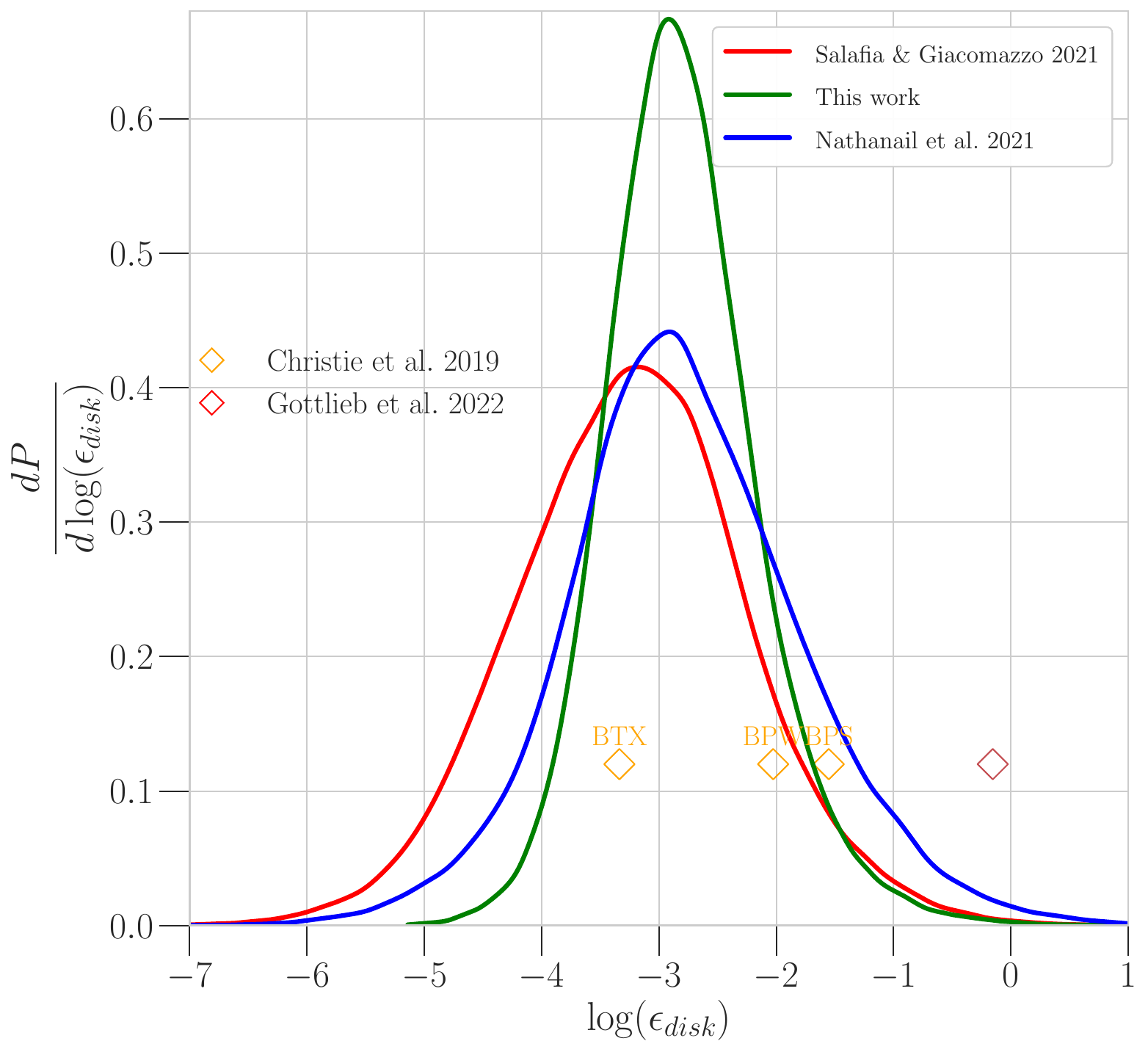}
        \includegraphics[width=.46\textwidth]{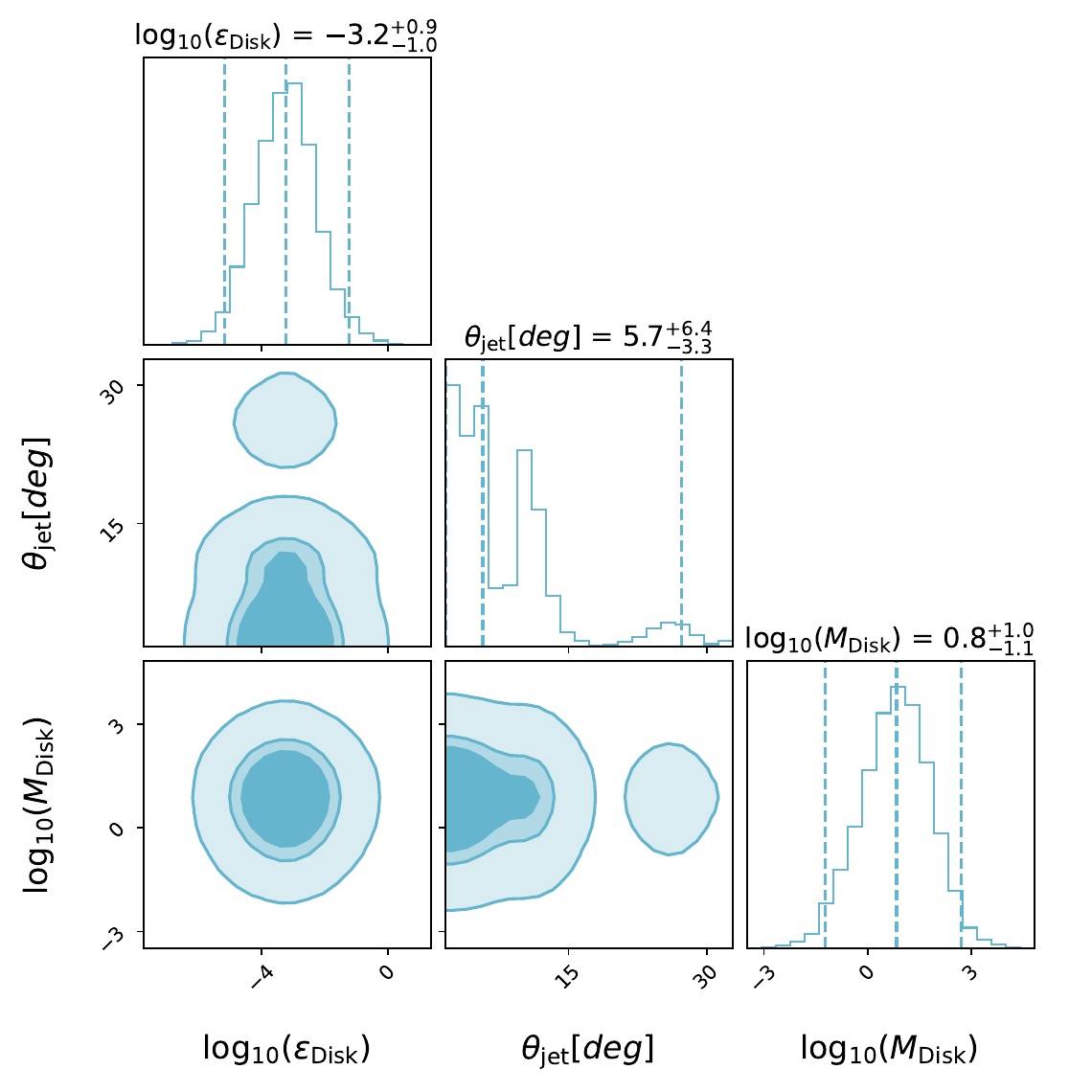}
        \caption{Left panel:
        Posteriors for the accretion to jet efficiency for
        the case of GRB170817A. The green line represent the efficiency derived
        from the algorithm described in this paper and the red line
        the one from \cite{2021A&A...645A..93S}, both of these make use of
        the kinetic energy distribution  from
        \citep{Ghirlanda2019}, whereas the blue line corresponds to the
        kinetic energy distribution from a strongly magnetized jet from
        \cite{2021MNRAS.502.1843N}. Over-plotted points refer to efficiencies
         from general relativistic magnetohydrodynamics simulations.
         Right panel:
         The resulting posterior distributions analyzing
         GRB201221D with the algorithm of this paper assuming an efficiency
	 similar to GRB170817A.}
\label{eff}
\end{figure*}

Jet-breakout time ($t_{\mathrm{jb}}$): We consider a jet with constant power
that is launched inside an  ejecta envelope with a power-law density profile. We follow the uncollimated approximation, presented by
\cite{2011ApJ...740..100B}. For more details see Appendix \ref{ApA}

Accretion-to-jet efficiency ($\epsilon _{\mathrm{disk}}$): We consider a
posterior distribution for this efficiency. To model this distribution, we examine two cases.
First, we derive the distribution for the accretion-to-jet efficiency,
derived from \cite{2021A&A...645A..93S}, which utilized the kinetic
energy distribution from \cite{Ghirlanda2019}.  Secondly, following the same methodology, we calculate the
posterior distribution for the accretion-to-jet efficiency  using a
different kinetic energy distribution. More specifically, we used the
energy profile that aligns with afterglow observations of GRB170817A,
assuming a strongly magnetized jet \citep{2021MNRAS.502.1843N}. Since the BNS merger systems are  qualitatively similar, we generalize the efficiency results for all sGRB cases.

% These
% guidelines are guided by the qualitative similarity observed in the
% evolution of BNS merger systems, justifying the reasonableness of our
% assumptions.

%By combining these approaches, we achieve a comprehensive
%understanding of the accretion-to-jet efficiency, enabling us to more
%effectively model and analyze the behavior of binary neutron star mergers
%and their associated gamma-ray bursts.

Total disk mass upon merger ($M_{\mathrm{disk}}$): We determine the total
disk mass upon merger using fitting formulas derived from numerical
simulations. 
These formulas primarily depend on the masses and tidal
deformabilities of the neutron stars \citep{ 2018ApJ...860...64F,
Radice_2018, Kruger2020, Barbieri2021}. For instance, in the case of
GW170817, LIGO provided posteriors for the tidal deformability and mass
of the binary components, allowing us to parametrically estimate the
remnant mass. Additionally, a portion of the total mass is extracted due
to the ejection mechanism before collapse, following the profile provided
in the Appendix A of \cite{Barbieri2021}, which can be summarized as follows:

\begin{equation}
M_{d2} = \dfrac{1}{4} (2+x_2) (x_2 - 1)^2 M_2,
\end{equation}

\begin{equation}
x_2 = 2[(1+\dfrac{M_1}{M_2})^{-1} + \lambda _2 ^{-1} -1],
\end{equation}

\begin{equation}
\lambda _2  = \bigg( \dfrac{M_2}{M_1}\bigg) ^{\beta} \bigg( \dfrac{\tilde{\Lambda}}{\Lambda_0}\bigg) ^{\alpha},
\end{equation}
where  $M_1$,$M_2$, $ \tilde{\Lambda}$ are the mass of the primary star, mass
of  the secondary star and  the dimensionless tidal deformability parameter of
the binary, with parameter  values of $\alpha =0.097$, $\beta = 0.241$ and $\Lambda _0 = 245$. The index
`1` of each quantity is calculated after interchanging `1` with `2`. Then, the disk mass of
the system after the merger is calculated as:

\begin{equation}
M_\mathrm{disk} = M_{d1}+M_{d2}
\end{equation}

Jet opening angle ($\theta_{\mathrm{jet}}$): We utilize a profile
constructed from the observations of a larger sample of short GRBs
\citep{2022arXiv221005695R}, to determine the distribution of the jet's
opening angle. This approach allows us to generalize our results to both
past and future short GRB candidates. Notably, this analysis reveals a
double peak in the jet's angle distribution at approximately 5 and 15
degrees. For the specific case of GRB170817A, an estimated opening angle
of 5-6 degrees 
\citep{Ghirlanda2019, 2019MNRAS.489.1919T, Mooley2018} or  ~ 15 degrees if a strongly magnetized
jet is assumed\citep{2021MNRAS.502.1843N}.  

Engine time ($t_{\mathrm{eng}}$): We assume that the jet is 
launched at the time of collapse. Gamma-ray emission begins after the
jet breaks out of the ejecta, and shuts down  upon the jet's ceasing (ignoring the
remaining jet's travel time, which reflects the time the jet spends
inside the ejecta, while the engine is turned off).  Consequently, we
associate the engine time with the sum of the observed quantity  $T_{90}$
and the jet break-out time ($t_{\mathrm{jb}}$).  

For the calculation of dynamical quantities, we use the following
relations:

Isotropic-equivalent jet's power ($L_{\mathrm{jet,iso}} $):
\begin{equation}
L_{\mathrm{GRB,iso}} = \epsilon_{\mathrm{GRB}} L_{\mathrm{jet,iso}} 
\end{equation}

Jet's opening angle ($\theta_{\mathrm{jet}}$):
\begin{equation}
\theta_{\mathrm{jet}} ^2 = \dfrac{L_{\mathrm{jet}} }{\pi  L_{\mathrm{jet,iso}} }
\label{t9s0}
\end{equation}

Emission ($T_{90}$) and gamma ray burst($t_{\mathrm{GRB}}$) time: 
\begin{equation}
T_{90} = t_{\mathrm{GRB}} =  t_{\mathrm{eng}} - t_{\mathrm{jb}}
\label{t90}
\end{equation}

Mass the jet must penetrate ($M_{\mathrm{ej}}$):
\begin{equation}
    M_{\mathrm{ej}} = M_{\mathrm{blue}}({t_\mathrm{coll}})
\label{mjet}
\end{equation}   %$t_{\mathrm{coll}}$

Effective disk mass ($M_{\mathrm{disk,eff}}$):
\begin{equation}
    M_{\mathrm{disk,eff}}  = M_{\mathrm{disk}} 
  - M_{\mathrm{blue}}(t_{\mathrm{coll}})
  - M_{\mathrm{red}}(t_{\mathrm{coll}})
\label{diskeff}
\end{equation}

Energy available for the jet ($E_{\mathrm{jet}}$):
\begin{equation}
E_{\mathrm{jet}} = \epsilon_{\mathrm{disk}} M_{\mathrm{disk,eff}} c^2 
\label{jet_ef}
\end{equation}

Total gamma-ray energy emitted ($E_{\mathrm{GRB}}$):
\begin{equation}
E_{\mathrm{GRB}} = \epsilon_{\mathrm{GRB}}E_{\mathrm{jet}}-\epsilon_{\mathrm{jb}}
  (L_{\mathrm{jet,iso}},t_{\mathrm{coll}}) E_{jb} 
\label{egrb}
\end{equation}

where $\epsilon _{jb}$ is the percentage of energy lost during the break-out
of the jet.
By definition, $\epsilon_{jb} =  E_{\mathrm{cocoon}}/L_\mathrm{jet}t_\mathrm{jet}$,where,
$E_{\mathrm{cocoon}} = (t_\mathrm{jb}-r_\mathrm{jb}/c)L_\mathrm{jet}$, and $r_\mathrm{jb}$ is the jet break-out radius. Therefore, from the energy released by the engine 
during break-out, equal to $E_{jb} = t_\mathrm{jb} L_\mathrm{jet}$ , only a portion is lost. This fraction is displayed in the lower panel of  Fig. \ref{JJ}.\footnote{The efficiency parameter scales with $t_{\mathrm{jb}} - r_{\mathrm{jb}}/c$. Keeping the same ejecta mass,
but increasing the velocity, will also lead to greater losses, but for
the sake of simplicity, we assume that velocity profiles do not  vary
significantly. }

What is yet not clarified by the above equations is the amount of mass
ejected and observed as either red or blue kilonova. These components of
the ejected mass are denoted as $M_{\mathrm{blue}}$ and
$M_{\mathrm{red}}$ and are calculated using the analytical formulas
presented in \cite{2019ApJ...876..139G}. Importantly, their determination
relies solely on the parameter $t_{\mathrm{coll}}$, which is the collapse
time of a super-massive neutron star to a black hole. The portion of the
mass denoted as $M_{\mathrm{red}}$ is mostly concentrated in large angles
towards the equator, and comes mostly from the dynamical ejecta and partially from 
the disk before the remnant collapses \citep{Bovard2017}. Thus, the mass
that the jet has to travel through is  the $M_{\mathrm{blue}}$
component (see Eq.  \eqref{mjet}). 

The set of equations outlined above serves a dual purpose. One 
is to utilize them to derive an $M_{\mathrm{disk}}$ value by analyzing
GW170817, similar to the methodology employed by
\cite{2021A&A...645A..93S}. It also allows us to derive the
efficiency of the accretion-to-jet energy conversion.  

\section{Applications}
\label{results}

\subsection{Jet efficiency} 

\begin{figure*}
\centering
\includegraphics[width=.47\textwidth]{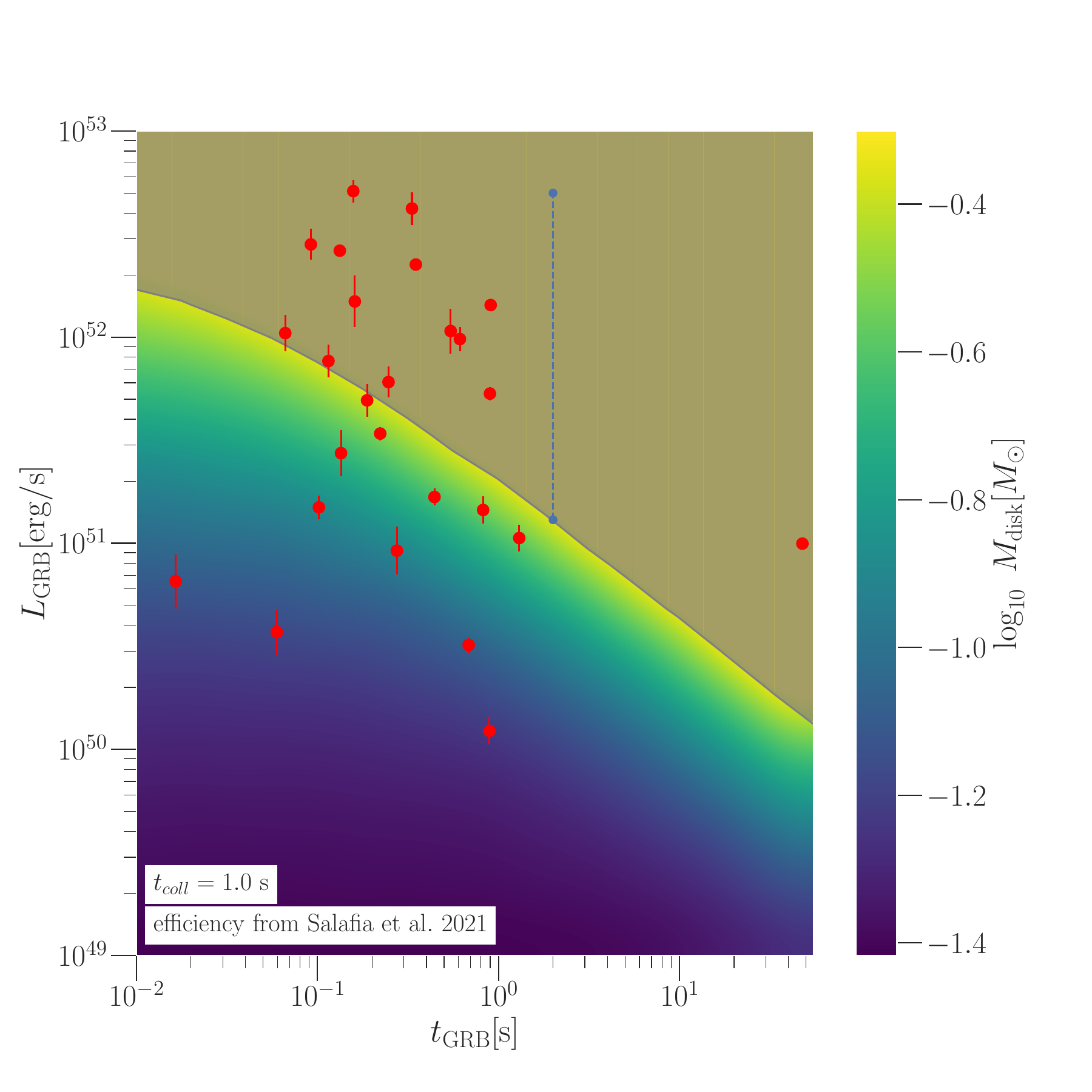}
\includegraphics[width=.47\textwidth]{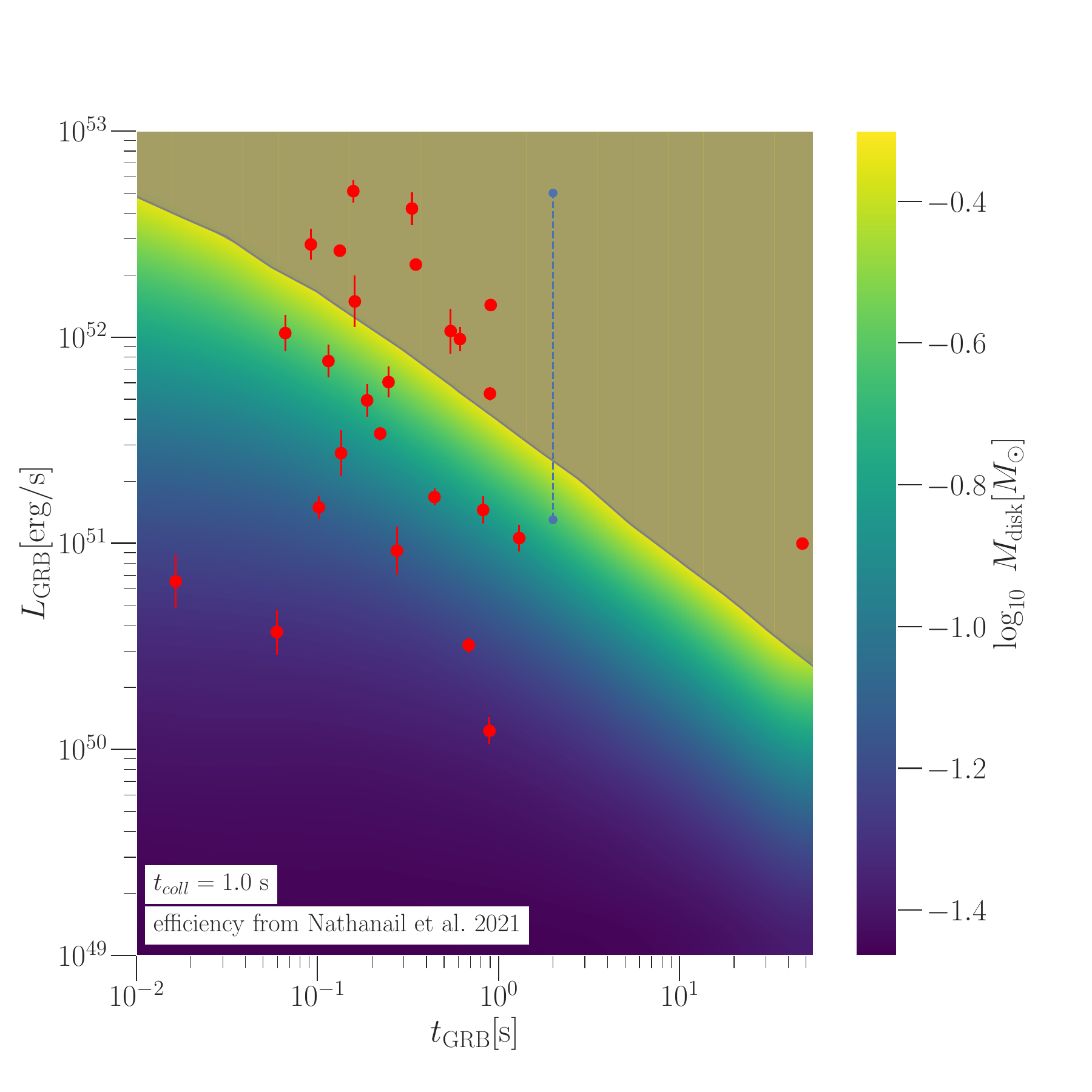}
        \caption{
        In both cases the background color represents the mass of the disk,
	calculated by the mean value of our posteriors.
        The red points represent observational events
        (refer to the main text for details). The
        blue vertical dashed line corresponds to the core luminosity of 
	GRB170817A. The shaded region indicates  where the
        disk mass exceeds $0.3  M_{\odot}$, which is an approximate
        limit derived from numerical relativity simulations. The solid grey line,
	marks the boundary between these two regions.  In the left
        plot,we used the efficiency profile from
        \citep{2021A&A...645A..93S}, while in the right, we adopted the
	corresponding efficiency when assuming a jet profile from
        \citep{2021MNRAS.502.1843N}.}
        \label{fig:panel}
\end{figure*}

Firstly, we recalculate the distribution of  accretion-to-jet efficiency, for the specific case of
GW170817, solving for $\epsilon_{\mathrm{disk}}$ in Eq. \ref{jet_ef}. With a Monte Carlo simulation, we draw the dimensionless tidal deformability parameter $\tilde{\Lambda}$ from \cite{2017PhRvL.119p1101A}. Based on the formula form \cite{Barbieri2021}, we calculate $M_{\text{disk}}(\tilde{\Lambda})$. Then, we calculate $M_{\text{disk,eff}}$ as $(1-f_w)M_{\text{disk}}$ with $f_w = 0.4$, according to \cite{2021A&A...645A..93S} (for the red curve only), or by Eq. \ref{diskeff} with $t_{\mathrm{coll}}=1$. For the jet's energy deposit, we simply draw from the posterior presented in either  \cite{Ghirlanda2019} or \cite{2021MNRAS.502.1843N}. Lastly, a trivial calculation gives us $\epsilon_{\mathrm{disk}}$. A summary is shown  in Table \ref{table:probability_distributions}.
The results for the efficiency of converting accretion energy 
into jet energy are illustrated in left panel of  
Fig. \ref{eff} and are represented by
the green line. They are comparable to the outcomes obtained by
\cite{2021A&A...645A..93S}, indicated by red in the same figure.
Furthermore , we examined the efficiency for a kinetic energy distribution
based on 3D general relativistic magnetohydrodynamics (GRMHD) simulations
conducted by \cite{2021MNRAS.502.1843N}. However, the analytical solution
we follow does not include magnetization effects,   and we treat each
kinetic distribution hydrodynamically.  Notably, the resulting
efficiency, when considering the latter kinetic energy distribution,
exhibited a slightly higher mean value.

\begin{table}[h!]
\centering
\caption{Probability Distributions for Various Quantities}    
\scriptsize % Further reduce the font size
\begin{tabular}{|c|c|}
\hline
\textbf{Quantity} & \textbf{Probability Distribution} \\
\hline
\( \tilde{\Lambda} \) & Distribution from \cite{2017PhRvL.119p1101A} \\
\hline
\( M_{\text{disk}} \) & \( M_{\text{disk}}(\tilde{\Lambda}) \) from \cite{Barbieri2021} \\
\hline
\( E_{\text{jet}} \) & posterior from \cite{Ghirlanda2019} / \cite{2021MNRAS.502.1843N} \\
\hline
\( \epsilon_\mathrm{disk} \) & Fig. \ref{eff} (from Eq. \ref{jet_ef}) \\
\hline
\end{tabular}
\label{table:probability_distributions}
\end{table}

Since GRB170817A was observed off-axis, the measured luminosity cannot be
directly utilized in our analysis. To address this, we need to account
for the isotropic luminosity as if the observation were on-axis.
\cite{Ghirlanda2019} argue that if this jet were observed directly along
its axis, its gamma-ray emission would have displayed an isotropic
equivalent luminosity of at least $L_{\mathrm{GRB,iso}} = 1 (\pm 0.35)
\times 10^{51} \mathrm{erg \ s^{-1}}$, assuming a $10\%$ efficiency in
converting kinetic energy to radiation and attributing a $35\%$ typical
error. Another assumption for GRB170817A concerns the association between
$T_{90}$ and $t_{\mathrm{GRB}}$. While this connection is evident for
on-axis observations, the dependence of the observed duration on the
viewing angle is not well understood. Note that these points have to be
revised if any observation of an on-axis short GRB accompanying a BNS GW
event is observed in the future.

Regarding the collapse time, we rely on the findings of
\cite{2019ApJ...876..139G}. Their methodology involves determining the
survival time of the merger remnant by integrating two distinct
constraints. Firstly, they calculate the time required for the generation
of the requisite mass of blue ejecta.  Simultaneously, they account for
the duration necessary for the relativistic jet to penetrate its way
through the expanding ejecta.  Through this dual constraint approach, it
is established that the remnant resulting from GW170817 must have
transitioned into a black hole after a collapse time of $t_{\rm coll} =
0.98_{-0.26}^{+0.31}$ seconds. An alternative interpretation of the delay
time, the time difference between the GW detection and  the onset of a
GRB, is presented in \cite{Beniamini2020}.

\subsection{Disk mass distribution for short GRBs} 

Our next objective is to draw more generalized conclusions about the disk
mass distribution  for observed short GRBs and argue if the obtained
result can be within the allowed limits for a BNS merger event. 
 
We systematically explore the parameter space encompassing $T_{90}$ and
$L_{\mathrm{GRB,iso}}$, requiring the associated disk mass necessary to
generate a GRB event corresponding to a specific point within this
parameter space. Details on how we make use of short GRBs and
their observed parameters throughout the algorithm, can be found in Appendix \ref{ApB}. Under the
assumption that the efficiency of accretion-to-jet efficiency is universal to
BNS merger events, we use the posterior distribution obtained in the
previous section, specifically based on the characteristics of GRB170817A. 
To comprehensively examine the influence of the
collapse time, we consider several scenarios: ranging from a duration of
$10^{-2} \mathrm{s}$ to $9 \mathrm{s}$. 

For each combination of ($t_{\mathrm{GRB}}$, $L_{\mathrm{GRB,iso}}$), we
generate a mass distribution by running a Monte Carlo simulation  for a
total of one million samples. To ensure a smoother dataset, we
incorporate an additional 100 data points for each draw, following a
normalized distribution with a $\sigma$ value equal to that sample's
variance. An illustrative example of the resulting distribution is
presented in the right panel of  Figure \ref{eff} for the case of 
GRB201221D. 

Our primary interest lies in determining an upper limit for the disk mass
for each combination $T_{90}$, $L_{\mathrm{jet}}$ in the parameter space,
which is what is observed from a regular short GRB with known distance.
We focus on the most probable value of the disk mass and assess its
feasibility within the context of BNS mergers.  In
Figure \ref{fig:panel}, we present the results of this analysis using a
color-scale representation. The solid grey line within the plot defines the
region where the disk mass aligns with the approximate maximum derived
from state-of-the-art numerical relativity simulations. The estimated
maximum is less than $0.3 M_{\odot}$, as reported in previous studies
\citep{Radice_2018, Kruger2020, Nedora2021,
Barbieri2021}. To ensure a conservative upper limit for the disk mass, we
set it at $0.3 M_{\odot}$. Notably, even for binary systems with a total
mass of approximately $3.3 M_{\odot}$ and significant asymmetry, which
generally leads to higher disk masses, the maximum disk mass
remains limited to around $0.1 M_{\odot}$ \citep{Camilletti2022}. Shaded,
olive color, areas on the plot represent regions with higher disk mass
values and essentially delineate the areas within the $T_{90}$,
$L_{\mathrm{jet}}$ parameter space where observational events are
statistically unlikely to be progenitors of BNS mergers, for the chosen efficiency values.

%It's important to note that the results presented in Figure
%\ref{fig:panel} are based on the underlying assumption that the %collapse
%time, $t_{\mathrm{coll}}$, is approximately $1 \mathrm{s}$."

 In our model, the ejecta's mass is controlled by collapse time (see Eq. \ref{mjet}).
Each panel sets the collapse time constant, and therefore the ejecta's mass.
An examination of horizontal line, sets constant $L_{\mathrm{GRB,iso}}$.
Therefore  the jet break-out time remains also a constant - see Fig A1 -. 
Consequently, the engine time can be rewritten as $T_{90}
+ t_{\mathrm{jb}}$ for these points. When the
engine time decreases, the merger's ejecta
consumes a larger portion of the disk's energy deposit. Therefore, the region where the collapse
time significantly influences the disk mass posterior primarily lies in
the lower $T_{90}$ range, where $T_\mathrm{eng} \gtrsim
t_{\mathrm{jb}}(L_{\mathrm{GRB,iso}},t_{\mathrm{coll}})$. 

However, it becomes evident that an even more crucial parameter is the
accretion-to-jet efficiency, which can span orders of magnitude.  One
approach to allow the shaded region to encompass disk masses compatible
with short GRBs from BNS mergers, is to allow the efficiency to vary
across the $(T_{90},L_{\mathrm{GRB,iso}})$ parameter space.  However this
is not easily visualized in a plot like Fig.\ref{fig:panel}, and  can
be better understood from the discussion for Table \ref{chance} in the
Appendix \ref{ApB}.

%----------------------------------------

\section{Conclusions}
\label{conc}

In this study, we developed a comprehensive algorithm to estimate the
dynamical quantities involved in short gamma-ray burst (GRB) events, with
a specific focus on the mass of the accretion disk formed after a 
BNS merger. Our approach linked observational parameters,
such as isotropic GRB luminosity ($L_{\mathrm{GRB,iso}}$) and burst
duration ($T_{90}$), to the properties of the merger remnant and its
ability to power a GRB event.

Our results highlight the significance of the jet efficiency and opening
angle in determining the disk mass required for a GRB event. The analysis
indicates that, in the parameter space of $T_{90}$ and
$L_{\mathrm{GRB,iso}}$, the majority of observational data correspond to
disk masses near $0.1$ solar masses ($M_{\odot}$), which is consistent with
current simulations. Importantly, some short GRB events exhibit a
significantly higher disk mass, raising questions about their origin.
This suggests the possibility that BNS mergers may involve different
mechanisms for jet launching than the well-studied GRB170817A, indicating
a potential lack of universality in the underlying physics. Specifically, if one 
changes the  efficiency parameter $\epsilon_{\mathrm{disk}}$, the observables that require high 
disk masses, will greatly vary. Future observations of 
sGRBs that will allow the calculation of this efficiency, will  verify or reject this assumption.

With the advancement of numerical relativity, we believe that these
methods can draw general restrictions with better confidence.
%The advancement of numerical relativity will also enable better
%prescription for the ejecta profiling and dependence on the collapse
%time. We expect that with methods described before, general restrictions
%can be drawn with better confidence.

----------------------------------------------------
\section*{Acknowledgements}
%-----------------------------------------------------------------------

The authors would like to thank P. Singh, R. Gill, S.I. Stathopoulosand L. Rezzolla   for
useful discussions.  Support also comes from the ERC Advanced Grant
``JETSET: Launching, propagation and emission of relativistic jets from
binary mergers and across mass scales'' (Grant No. 884631).
%The simulations were performed on SuperMUC at LRZ in Garching, on the
%GOETHE-HLR cluster at CSC in Frankfurt, and on the HPE Apollo Hawk at
%the High Performance Computing Center Stuttgart (HLRS) under the grant
%numbers BBHDISKS and BNSMIC.

%----------------------------------------------------------------------
\section*{Data Availability}
%-----------------------------------------------------------------------
%\smallskip\noindent\textit{Data Availability.~}
The data underlying this article will be shared on reasonable request to the
corresponding author.

--------------------------------------------------------

\bibliographystyle{aa}%{aasjournal}    %{mnras}
\bibliography{sgrb}

\appendix %First appendix

\section{Jet's evolution}\label{ApA}

We solve for the dynamics of an arbitrary jet, following the
uncollimated case of \cite{2011ApJ...740..100B}. We assume  an identical
prescription for the rest of our work, regarding   the jet's dynamical
evolution. We will briefly describe the governing equations

For simplicity, we consider a  jet with constant power, that is launched
inside an ejecta envelope. The ejecta envelope consists of mass ejected
through various channels. The main mechanisms can be recognized as the
dynamical ejection \citep{2018ApJS..234....1B},  the neutrino-driven
winds\citep{2015ApJ...813....2M}, and the magnetically-driven
winds\citep{2018ApJ...860...64F}. Ejecta  reach semi-relativistic
velocities and they also act as a "barrier" that the jet has to drill
through  before reaching the ISM.

The velocity profile is the following:
\begin{equation}
   v_{\mathrm{ej}} = 0.3 c \dfrac{r}{r_{\mathrm{out}}}
   \label{uu}
\end{equation}
which is close to numerical values \citep{2019ARNPS..69...41S}  and
observational constraints \citep{2017Natur.551...64A, 2017Sci...358.1570D,
2017Sci...358.1574S, 2017Natur.551...67P, 2017Sci...358.1559K,
2017Natur.551...80K, 2017Sci...358.1565E, 2017Natur.551...75S}.

For the ejecta density profile, we adopt a simple power-law

\begin{equation}
\rho(r,t) = \dfrac{1}{4\pi}\dfrac{M_{ej}(t)}{r_{\mathrm{out,0}^3}}
	\bigg[\dfrac{r}{r_\mathrm{out}(t)}\bigg]^{-2}
\label{bo}
\end{equation}

where $r_\mathrm{out}(t)$ is the outer radius of the ejecta envelope,
and is moving accordingly as the ejecta moves outwards. $t$ is the time
passed after the collapse. The radius is calculated via 

\begin{equation}r_\mathrm{out}(t) = 0.3 \ (t+t_{\mathrm{coll}})\ c +
r_{\mathrm{out,0}}\end{equation}

where $r_{\mathrm{out,0}} = 3   \times 10^{6} \mathrm{cm}$.
$M_{\mathrm{ej}}(t)$ denotes the mass of the blue component of the ejecta,
which is produced by the BNS compact remnant, 
before collapsing to a black hole at $t_{\mathrm{col}}$. Its value is
calculated from the formula reported in \cite{2019ApJ...876..139G}.

Assuming that the jet injection starts at the collapse time, we
analytically find the velocity of the jet's head, which is slowed down by
the double shock that is ignited upon collision with the ejecta.  The
jet's head velocity, while inside the ejecta envelope, is dictated by the
ram pressure equilibrium in the head's frame.

\begin{figure}
\centering
\includegraphics[width=.44\textwidth]{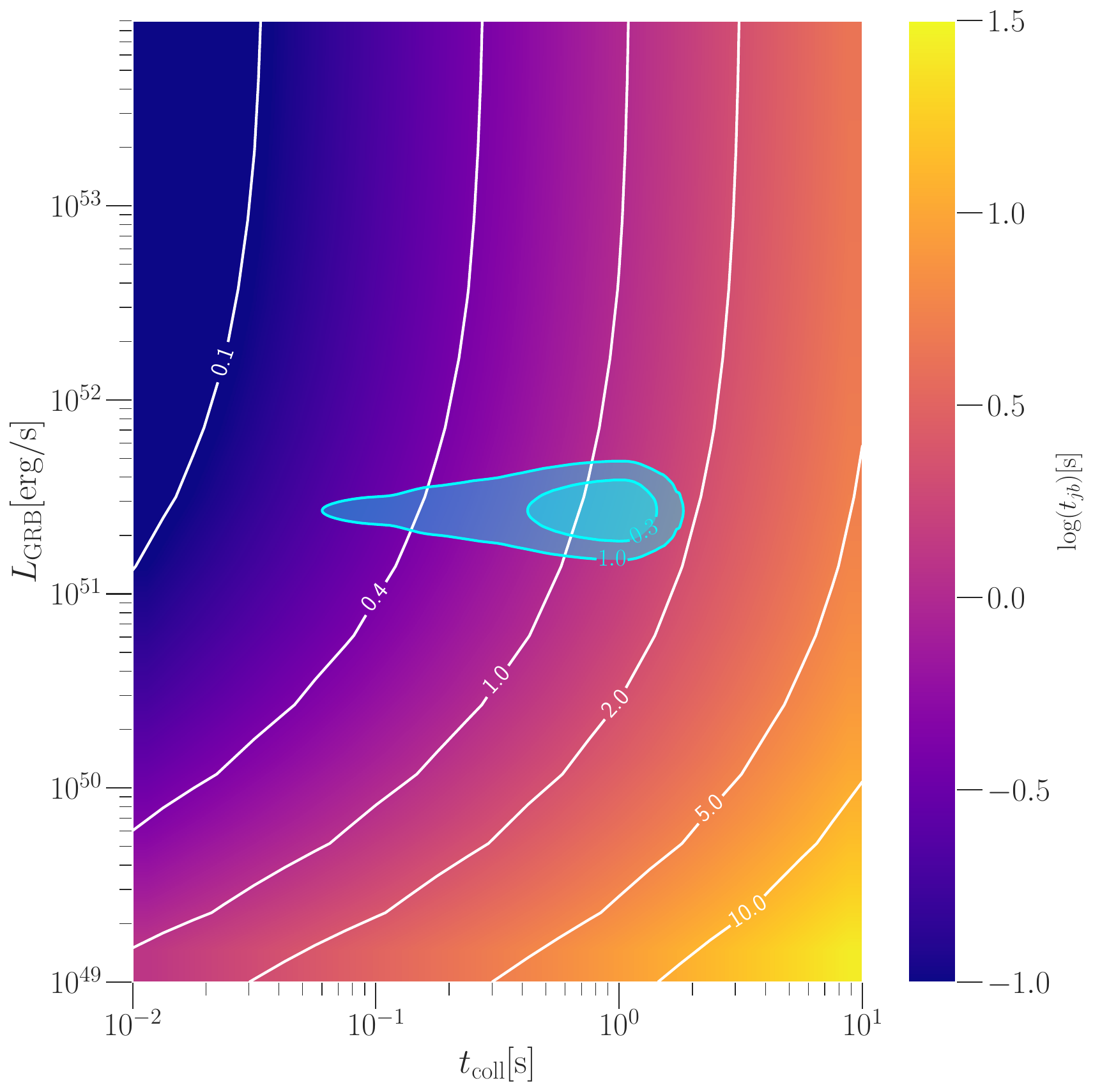}
\includegraphics[width=.44\textwidth]{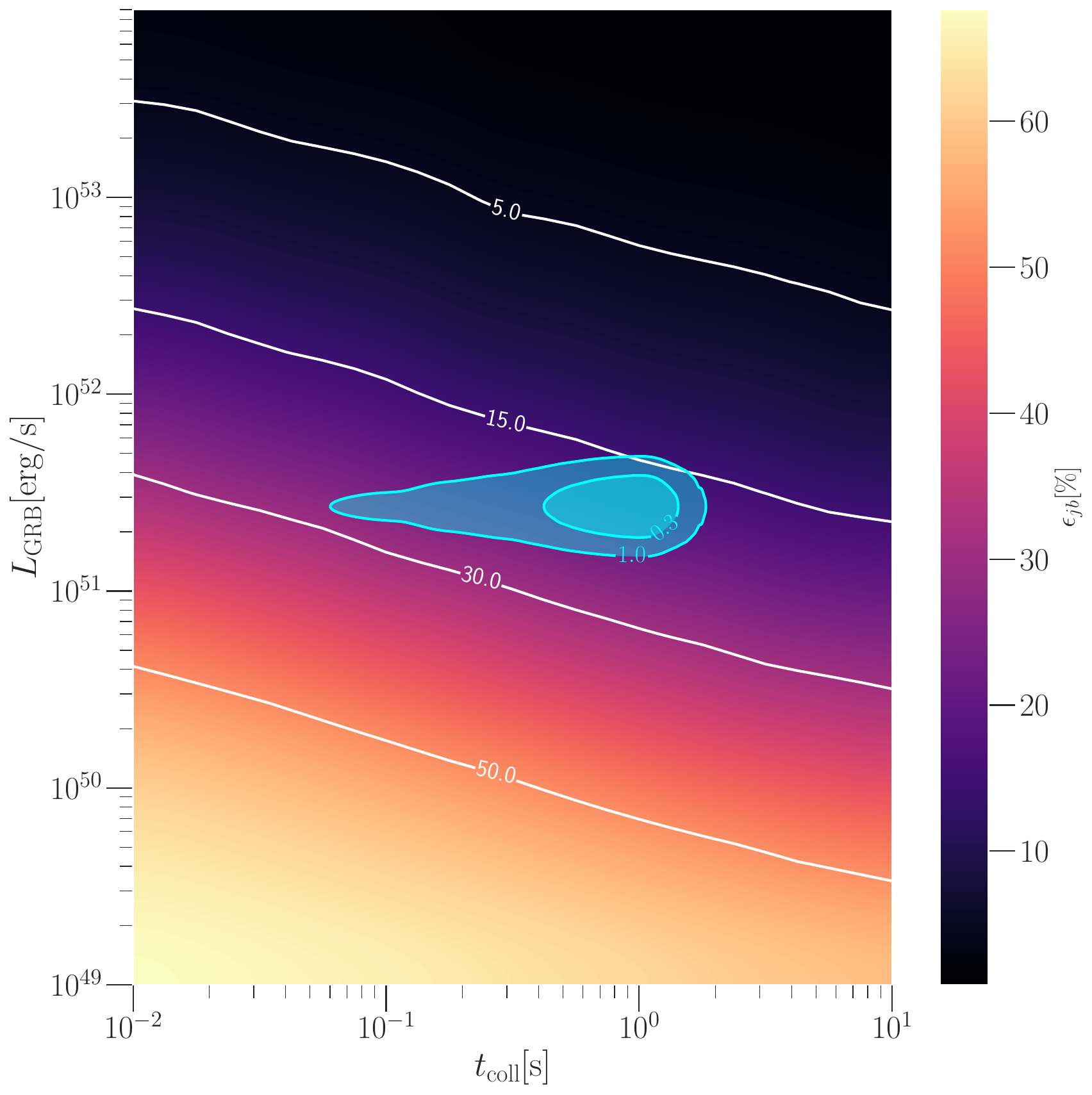}
\caption{
Upper panel: jet break-out time (in
color) for the uncollimated case. Lower panel: the fraction of the jet
energy lost for break-out $\epsilon _{\mathrm{jb}}$ (in color).
Quantities in both panels are plotted as functions of
$t_{\mathrm{coll}}$ and $L_{\mathrm{GRB,iso}}$. The cyan region
corresponds to 1 and 2$-\sigma$ estimations of GRB170817A.}
\label{JJ}
\end{figure}

\begin{equation}
  \rho_{j}h_{j}[\Gamma_{j}\Gamma_{h}(\beta_{j}-\beta_{h})]^{2} =
  \rho_{ej}h_{ej}[\Gamma_{ej}\Gamma_{h}(\beta_{h}-\beta_{ej})]^{2}
\end{equation}

where  $\rho$ and $h$ are the mass density and specific enthalpy of each
fluid and $\beta$ the velocity. With the sub indices $j, \, h$ and $ej$
we denote the jet, the jet's head and the ejecta.  Assuming a
relativistic jet that penetrates through cold ejecta, it can be concluded
that:   % [eg]\citep{2003MNRAS.345..575M}:

\begin{equation}\beta_h = \dfrac{1+\Tilde{L}^{-0.5}
\beta_{ej}(r_h)}{1+\Tilde{L}^{-0.5}}\end{equation} 

where $r_h$ is the jet's head position and $\Tilde{L}$ is  the
ratio of the jet energy density, to the  ejecta density  at that
position.  

\begin{equation}\Tilde{L} \approx
\dfrac{L_j}{\Sigma_j \rho _{\mathrm{ej}}} c^3
\end{equation}

%Making assumptions for the cylindrical jet's geometry we get:

% \begin{equation}p_c = \dfrac{L_j}{3}\dfrac{\int(1-\beta_h)dt}{\big( \int \beta_c dt \big)^2
% z_h}\end{equation} and \begin{equation}\beta_c = \big[\dfrac{p_c}{\rho_{\mathrm{ej}}c^2}\big]^{0.5}\end{equation}

With given $t_{\mathrm{coll}},L_{\mathrm{jet,iso}}$, and the assumption
of uncollimated jet, the system described above can be solved for the jet
evolution.  The break-out time computed is shown on the upper panel of
Fig. \ref{JJ}, as a function of $t_{\mathrm{coll}}$ and
$L_{\mathrm{jet,iso}}$.  Under the assumption of quasi-spherical ejecta,
meaning that within the opening angle of the jet, the ejecta are
spherically symmetric, this quantity does not depend
on the opening angle of the jet.

The drilling  that the jet has to go through, reduces its available
energy which is dissipated later to produce the observed emission.  The
higher the resistance, meaning the more  massive the ejecta  to be
bypassed,  the greater the losses in energy before break out. The energy lost, is better
expressed as follows:

\begin{equation}
E_\mathrm{lost} = \epsilon_{jb} t_\mathrm{jb} L_{\mathrm{jet}}
\end{equation}

Notice the similarities between this formulation presented and  the one
developed in \cite{2019ApJ...876..139G} - see sections 5 and 6 -. In \cite{2019ApJ...876..139G} 
reference case is GRB170817A, which from a dynamical point of view,
there is no reason to consist an exception compared to the other sGRBs
examined in this work. Other simplified and intuitive approximations, as the one from
\cite{2018ApJ...866....3D}, where numerical results are employed, could not be adapted 
to our scheme, since in the latter for example, the ejecta radius is set to a constant upon 
jet launching, when in our case, it varies based on the collapse time, which dictates how much time 
the outer shell of the ejecta has, to freely expand. However, modifying the notation presented, to the
same parameters as in \cite{2018ApJ...866....3D}, the break-out time  differed mostly by an order of 
$\sim 2$, which can be explained by slightly modifying the constant parameters in the fitting procedure.

 %First appendix

\section{Observables}\label{ApB}

\begin{figure*} \centering
\includegraphics[width=.45\textwidth]{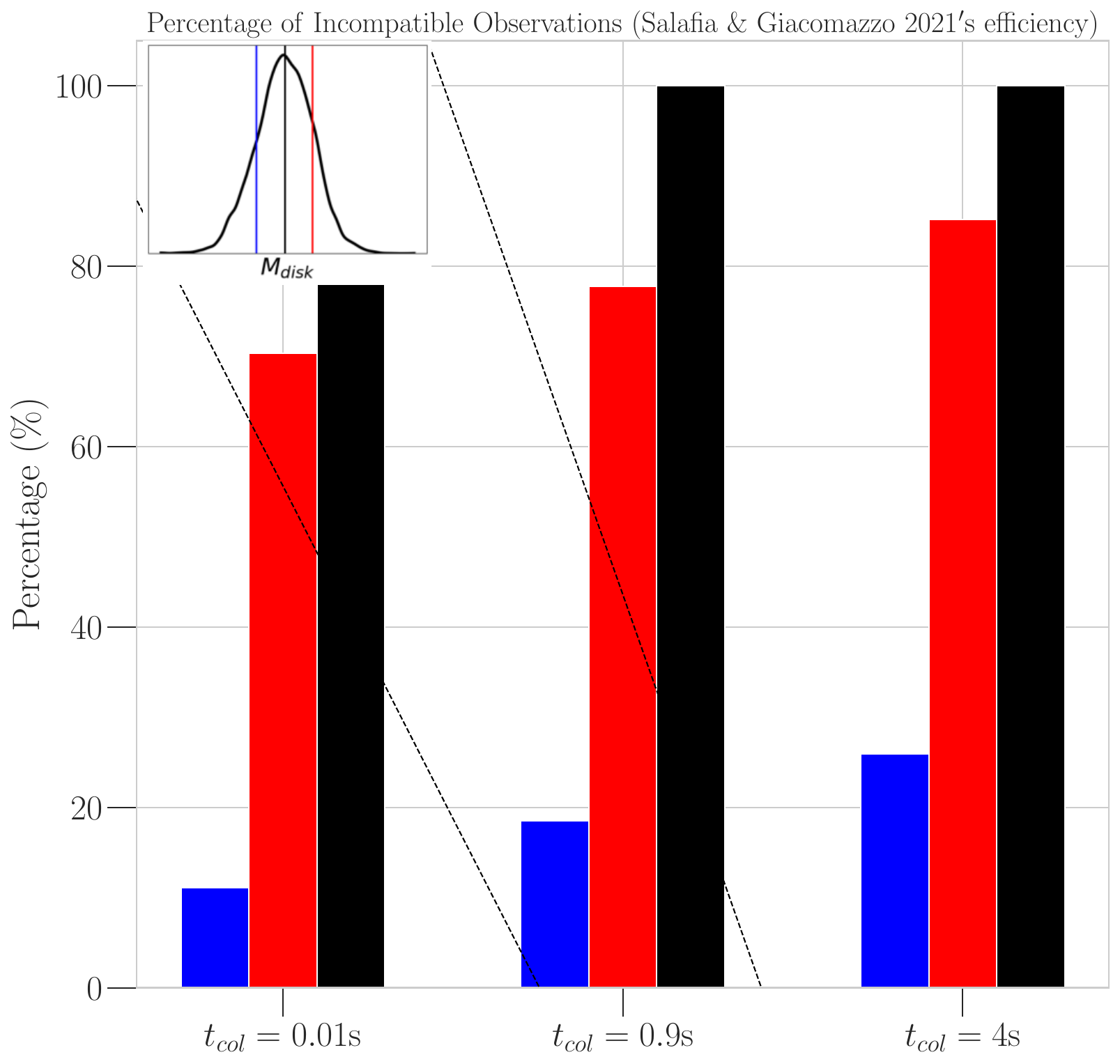}
\raisebox{-0.0cm}{\includegraphics[width=.45\textwidth]{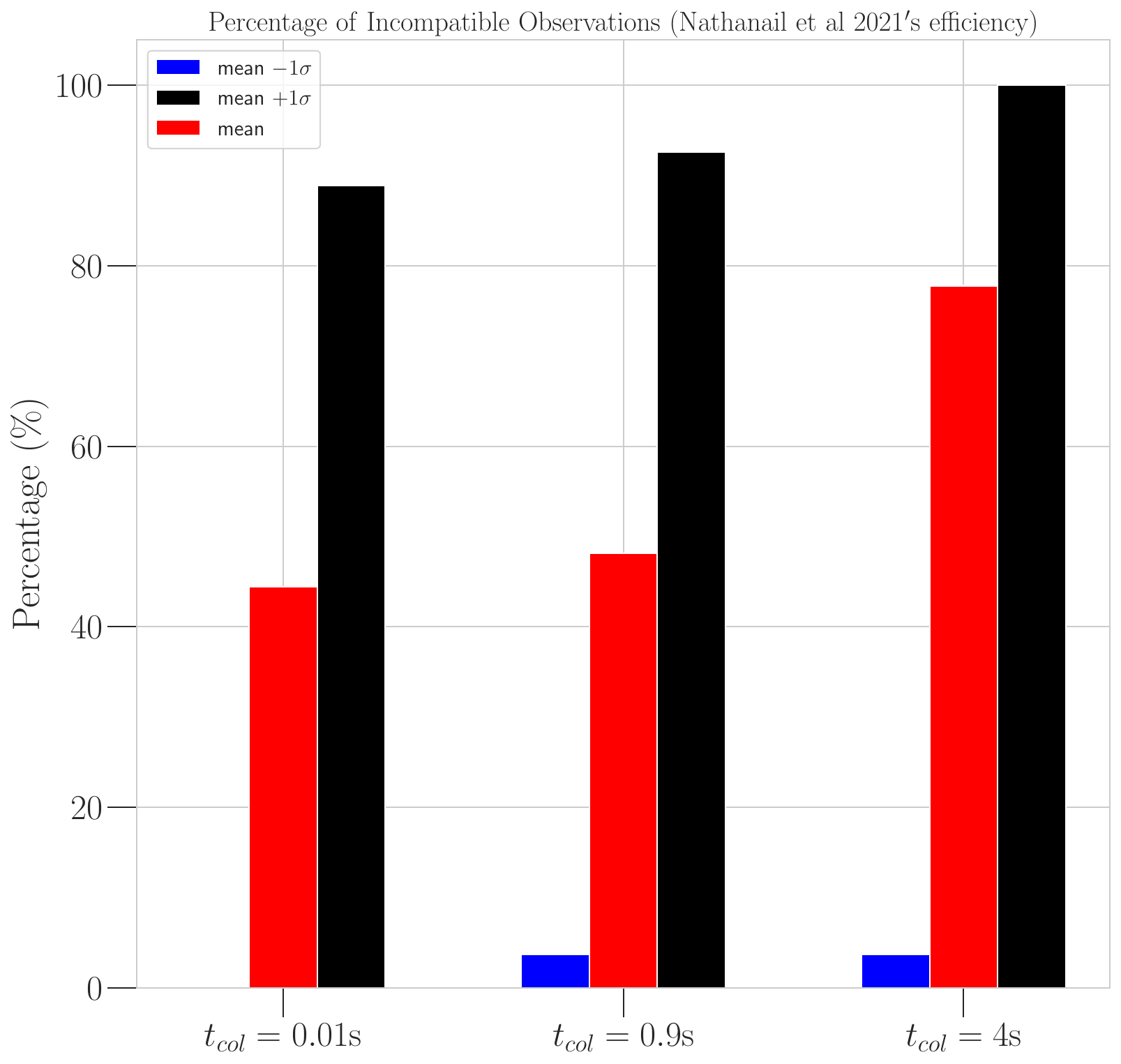}}
        \caption{
        Variation in the percentage of short GRBs data classified as
        incompatible with BNS-merger as a function of collapse time (see
        Fig. \ref{fig:panel}). The differently colored columns represent
        different levels of disk mass selection for each observation
        event: the ranges covering $33\%$, $50\%$, and $67\%$ of the disk
        mass posterior. The data reveals an increasing trend with higher
	collapse time.}
\label{cuts}
\end{figure*}

In this appendix, we thoroughly explore real observations of short GRBs within the
$(T_{90}, L_{\mathrm{GRB,iso}})$ parameter space as shown in Figure
\ref{fig:panel}. Our objective is to assess what percentage of these
observations can be reliably interpreted as a result of  
BNS mergers. To achieve this, we utilize publicly available data
from the GRB archive of the Neil Gehrels Swift Observatory
\citep{2004ApJ...611.1005G}. We specifically focus on short GRBs,
characterized by observed $T_{90}$ durations less than 2 seconds.
Additionally, we filter this subset to include only events with known
redshift measurements, which are essential for estimating the isotropic
$\gamma$-ray luminosity.

To calculate the isotropic $\gamma$-ray luminosity
($L_{\gamma,\mathrm{iso}}$), we use the BAT fluence ($\Phi$) along with
the luminosity distance ($d_L(z)$), redshift ($z$), and the Band function
\citep{1993ApJ...413..281B} parameters ($a = -0.5, b = -2.25$) to model
the differential photon spectrum within the 1 keV–10 MeV energy range. We
assume a rest-frame peak energy of $E_p = 800 \mathrm{keV}$
\citep{2011A&A...530A..21N}.

\begin{equation}
    L_{\gamma,\mathrm{iso}} = \dfrac{4\pi d_L(z) \Phi}{T_{90}}
  \dfrac{\int_{1\ \mathrm{keV}}^{10 MeV} dE EN(E)}{\int_{15(1+z)\
  \mathrm{keV}}^{150(1+z) keV} dE EN(E)}
  \label{apd:b}
\end{equation}

\begin{table}[h!]
\centering
\caption{ Sources for Various Quantities}
\small % Reduce the font size
\begin{tabular}{|c|c|}
\hline
\textbf{Quantity} & \textbf{Source} \\
\hline
\( L_{\gamma\mathrm{,iso}} \) & Observable quantity - Eq. \ref{apd:b} \\
\hline
\( \theta_{\mathrm{jet}} \) & Distribution from \citep{2022arXiv221005695R} \\
\hline
\( \epsilon_{\text{disk}} \) & Result from Fig 1 \\
\hline
\( \epsilon_{\text{grb}} \) & 0.15 \\
\hline
\( E_\mathrm{jet}\) & solution from Eq. \ref{egrb} \\
\hline
\( M_\mathrm{disk,eff}\) & solution from Eq. \ref{jet_ef} \\
\hline
\( M_\mathrm{disk}\) & solution from Eq. \ref{diskeff} \\
\hline
\( t_{\text{col}} \) & varying in x-axis \\
\hline
\end{tabular}
\label{table:sources}
\end{table}

Following the calculation of the isotropic $\gamma$-ray luminosity, and
maintaining the assumption of $t_{\mathrm{coll}}=1\mathrm{s}$. Our
algorithm generates a posterior distribution for the disk mass  by 
the following way. First, we select every source from the aforementioned subset, and 
calculate $L_{\mathrm{jet,iso}}$ from Eq. \ref{apd:b}. Then we draw disk to jet efficiency, 
from the distribution presented in Fig. \ref{eff}, and an opening angle from 
\citep{2022arXiv221005695R}. We can then calculate for each observation $L_{\mathrm{jet,iso}}$ 
-we use a constant $\epsilon_\mathrm{GRB}$-. We solve for $E_\mathrm{jet}$ from 
 Eq. \ref{egrb}. Then, a simple calculation from Eq. \ref{jet_ef} gives $M_\mathrm{disk,eff}$,
and since collapse time is constant, we employ Eq. \ref{diskeff}, to solve for $M_\mathrm{disk}$ . 
This distribution is exemplified in the right panel of Figure \ref{eff},
illustrating the specific case of GRB 211221D. From this distribution, and for each observation, we
derive the mean value and the $1$-$\sigma$ range. These short GRB events
are then mapped onto the $T_{90}$, $L_{\mathrm{jet}}$ parameter space, as
illustrated in Figure \ref{fig:panel}. We assess the compatibility of
each event with a BNS merger based on its position within the shaded
region. Notably, we include GRB 211211A, even though it belongs to the
category of long-duration GRBs, as it is classified as a burst
originating from a compact object merger, supported by kilonova
measurements and host property analysis \citep{2022Natur.612..228T}.

To quantify the percentage of incompatible cases, we use blue, red, and
black lines to represent $33\%$, $50\%$, and $67\%$ quantiles of the
posterior distribution. These lines correspond to the mean value minus
$1$-$\sigma$, the mean value, and the mean value plus $1$-$\sigma$,
respectively. The results of this analysis are displayed in Fig.
\ref{cuts}, where we also explore the influence of varying the collapse
time, extending from the expected $t_{\mathrm{coll}}=1\mathrm{s}$ for GRB
170817A to smaller and larger values.  The left panel of Fig. \ref{cuts}
assumes the efficiency derived by \cite{2021A&A...645A..93S} for
an empirical structured jet, while for the right one the energy
efficiency distribution is based on the kinetic
energy from \cite{2021MNRAS.502.1843N}.

The handling of each 
parameter is shown in Table \ref{table:sources}.
Fig. \ref{cuts} demonstrates that the variation in collapse time has a
rather limited impact on the results compared to other factors,  where
we followed the exact same procedure described earlier, but for different collapse times. Our
in-depth analysis highlights the greater significance of parameters like
efficiency and opening angle. However, we observed a positive correlation
between the percentage of possible BNS merger
events and shorter collapse times.

\begin{table}
    \centering
    \caption{Probability for BNS event  and  corresponding Disk Mass in $\log_{10}[M_\odot]$  with $t_\mathrm{col}=1\mathrm{sec}$}
    \begin{tabular}{|l|c|c|c|c|}
    \hline
    & \multicolumn{2}{c|}{Salafia et al. 2021} & \multicolumn{2}{c|}{Nathanail et al. 2021} \\
    \hline
	    GRB Name & P & $\log(M_{\mathrm{Disk}}[M_\odot])$ & P & $\log(M_{\mathrm{Disk}}[M_\odot])$ \\
    \hline
    201221D & 0.33 & $0.85_{-0.99}^{+1.04}$ & 0.44 & $0.56_{-0.98}^{+1.09}$ \\
    200522A & 0.79 & $-0.30_{-0.86}^{+0.78}$ & 0.87 & $-0.58_{-0.86}^{+0.66}$ \\
    190627A & 0.39 & $0.68_{-0.97}^{+1.03}$ & 0.53 & $0.33_{-1.05}^{+1.03}$ \\
    160624A & 0.81 & $-0.37_{-0.88}^{+0.76}$ & 0.89 & $-0.66_{-0.85}^{+0.60}$ \\
    150423A & 0.53 & $0.33_{-0.95}^{+0.97}$ & 0.63 & $0.06_{-0.98}^{+0.99}$ \\
    150120A & 0.74 & $-0.19_{-0.91}^{+0.86}$ & 0.83 & $-0.47_{-0.90}^{+0.74}$ \\
    150101B & 0.94 & $-0.75_{-0.75}^{+0.51}$ & 0.98 & $-0.96_{-0.67}^{+0.35}$ \\
    141212A & 0.72 & $-0.13_{-0.88}^{+0.84}$ & 0.82 & $-0.45_{-0.91}^{+0.73}$ \\
    140903A & 0.77 & $-0.23_{-0.86}^{+0.81}$ & 0.86 & $-0.54_{-0.88}^{+0.68}$ \\
    140622A & 0.70 & $-0.07_{-0.89}^{+0.86}$ & 0.81 & $-0.37_{-0.89}^{+0.78}$ \\
    131004A & 0.50 & $0.39_{-0.94}^{+1.00}$ & 0.64 & $0.04_{-1.01}^{+0.96}$ \\
    130603B & 0.54 & $0.30_{-0.92}^{+0.99}$ & 0.67 & $-0.04_{-0.97}^{+0.94}$ \\
    101219A & 0.43 & $0.59_{-0.95}^{+1.04}$ & 0.54 & $0.28_{-0.99}^{+1.03}$ \\
    100724A & 0.42 & $0.61_{-0.96}^{+1.05}$ & 0.55 & $0.27_{-1.01}^{+1.03}$ \\
    090510 & 0.40 & $0.68_{-0.99}^{+1.06}$ & 0.53 & $0.33_{-1.00}^{+1.04}$ \\
    090426 & 0.25 & $1.12_{-0.99}^{+1.08}$ & 0.36 & $0.79_{-1.01}^{+1.14}$ \\
    080905A & 0.97 & $-0.86_{-0.71}^{+0.43}$ & 0.99 & $-1.03_{-0.58}^{+0.29}$ \\
    071227 & 0.74 & $-0.18_{-0.90}^{+0.83}$ & 0.84 & $-0.46_{-0.88}^{+0.72}$ \\
    070724A & 0.87 & $-0.54_{-0.83}^{+0.66}$ & 0.93 & $-0.78_{-0.79}^{+0.51}$ \\
    070429B & 0.65 & $0.04_{-0.93}^{+0.91}$ & 0.76 & $-0.28_{-0.96}^{+0.85}$ \\
    061217 & 0.70 & $-0.09_{-0.92}^{+0.87}$ & 0.80 & $-0.37_{-0.93}^{+0.77}$ \\
    061201 & 0.94 & $-0.73_{-0.76}^{+0.53}$ & 0.96 & $-0.90_{-0.69}^{+0.41}$ \\
    060502B & 0.88 & $-0.56_{-0.83}^{+0.64}$ & 0.94 & $-0.79_{-0.75}^{+0.49}$ \\
    051221A & 0.37 & $0.72_{-0.95}^{+0.99}$ & 0.49 & $0.43_{-1.00}^{+1.10}$ \\
    050813 & 0.52 & $0.35_{-0.93}^{+1.01}$ & 0.65 & $0.02_{-0.99}^{+0.97}$ \\
    050509B & 0.96 & $-0.83_{-0.73}^{+0.45}$ & 0.98 & $-1.01_{-0.62}^{+0.31}$ \\
    211211A & 0.25 & $1.09_{-0.98}^{+1.05}$ & 0.36 & $0.78_{-1.00}^{+1.13}$ \\
    \hline
    \end{tabular}
    \label{chance}
\end{table}

For each short GRB event in the sample, while maintaining  $t_\mathrm{col}=1\mathrm{sec}$, we are investigating, we estimate
the probability of the event originating from a BNS merger, with a
specific focus on the mass of the disk surrounding the merger remnant.
The approach to assigning a probability is as follows: from the
normalized posterior distribution, which integrates to a sum of 1, we
compute the integral for the region with a lower mass than the
conservative upper limit for the disk mass, which is set to $0.3 M_{\odot}$.
This value represents the probability. These results are reported
in Table \ref{chance}. We provide the disk mass values calculated
based on the accretion-to-jet energy efficiency profiles from both
\cite{2021A&A...645A..93S} (left columns) and \cite{2021MNRAS.502.1843N}
(right columns).

\label{lastpage}
\end{document}